\documentclass[epj]{svjour}
\usepackage{epsfig}

\newcommand  {\rsq}    {\left< r^{2} \right>}

\renewcommand{\DH}{Debye--H\"uckel\ }
\newcommand{\OSFKK}{OSFKK\ }
\newcommand{\KMBJ}{KMBJ\ }

\begin{document}

\title{The Electrostatic Persistence Length of Polymers 
beyond the OSF Limit}
\titlerunning{The Electrostatic Persistence Length of Polymers 
beyond the OSF Limit}
\author{Ralf Everaers\inst{1}, Andrey Milchev\inst{1,2} and Vesselin Yamakov\inst{1,2}
\thanks{present address: Argonne National Laboratory,
  Materials Science Division, Build. 212,
  9700 S. Cass Avenue, Argonne, IL--60439, USA}
}
\authorrunning{R. Everaers, A. Milchev and V. Yamakov}

\institute{
Max--Planck--Institut f\"ur Polymerforschung,
  Postfach 3148, D--55021 Mainz, Germany \and
Institute for Physical Chemistry, Bulgarian Academy
  of Sciences, G. Bonchev Street, Block 11, 1113 Sofia, Bulgaria}

\abstract{ 
  We use large scale Monte Carlo simulations to test scaling theories
  for the electrostatic persistence length $l_e$ of isolated,
  uniformly charged polymers with \DH intrachain interactions in the
  limit where the screening length $\kappa^{-1}$ exceeds the intrinsic
  persistence length of the chains.  Our simulations cover a
  significantly larger part of the parameter space than previous
  studies.  We observe no significant deviations from the prediction
  $l_e\propto\kappa^{-2}$ by Khokhlov and Khachaturian which is based
  on applying the Odijk-Skolnick-Fixman theory to the stretched de
  Gennes-Pincus-Velasco-Brochard polyelectrolyte blob chain. A linear
  or sublinear dependence of the persistence length on the screening
  length can be ruled out. We argue that previous numerical results
  pointing into this direction are probably due to a combination of
  excluded volume and finite chain length effects.  The paper
  emphasizes the role of scaling arguments in the development of
  useful representations for experimental and simulation data.
}  

\maketitle
\section{Introduction}
The theoretical understanding of macromolecules carrying ionizable
groups is far from complete~\cite{BarratJoanny_acp_96,Ullner_handbook_02}. In spite of
the long range of the interactions, the systems are often discussed
using analogies to neutral polymers.  A prominent example is the
concept of an electrostatic persistence length, which was introduced
more than 20 years ago by Odijk~\cite{Odijk_jpspp_77} and by Skolnick
and Fixman~\cite{SkolnickFixman_mm_77} (OSF). They considered a
semiflexible polymer or wormlike chain (WLC) with intrinsic
persistence length $l_0$ and Debye-H\"uckel (DH) screened
electrostatic interactions $U_{DH}/k_BT = (q^2 l_B/r) \exp(-\kappa r)$
between charges $q/e$ spaced at regular intervals $A$ along the chain.
The Bjerrum length $\l_B$ characterizes the strength of the
electrostatic interactions and is defined as the distance where
the Coulomb energy of two unit charges $e$ is equal to $k_BT$.  Due to
the presence of mobile ions the bare Coulomb interaction is cut off
beyond the screening length $\kappa^{-1}$.  OSF were interested in
bending fluctuations and considered the resulting increase of the
electrostatic energy relative to the straight ground state.  In the
(``OSF'') limit $\kappa^{-1}\ll l_0$ where the screening length is
smaller than the intrinsic persistence length of the chain and to
lowest order in the local curvature, the \DH interaction makes an
additive contribution to the bending rigidity.  As a consequence, a
WLC with DH interactions (DHWLC) behaves in this limit on large length
scales like an ordinary WLC with renormalized persistence length

\begin{eqnarray}
l_p = l_0 + l_{OSF}\\
l_{OSF} = \frac{q^2 l_B}{4 A^2 \kappa^2}
\label{eq:OSF}
\end{eqnarray}
Ever since, there has been a lively debate on how to extend the theory
to parameter ranges beyond the OSF limit $\kappa^{-1}\ll l_0$.  Barrat
and Joanny (BJ)\cite{BarratJoanny_epl_93} have shown that the original
OSF derivation breaks down, if the chains start to bend significantly
on length scales comparable to the screening length.  As a
consequence, Eq.~(\ref{eq:OSF}) cannot simply remain valid beyond the
OSF limit as was sometimes speculated~\cite{Odijk_jpspp_78}.  Two main
scenarios, which we denote by ``\OSFKK'' and ``\KMBJ'' after the
initials of the main authors, have been discussed in the literature: 
\begin{description}
\item[\OSFKK] According to Khokhlov and Khachaturian
  (KK)~\cite{KhokhlovKhachaturian_pol_82} the OSF theory can be
  applied to a ``stretched chain of polyelectrolyte blobs'', a concept
  introduced by de Gennes et al.~\cite{GPVB_jp_76} to describe the
  behavior of weakly charged flexible polyelectrolytes in the absence
  of screening. The persistence length of the blob chain is then
  calculated from Eq.~(\ref{eq:OSF}) using suitably renormalized
  parameters.
\item[\KMBJ] Refinements by
  Muthukumar~\cite{Muthukumar_jcp_87,Muthukumar_jcp_96,Muthukumar_jcp_01}
  of the original theory of Katchalsky~\cite{Katchalsky_ahca_48} treat
  electrostatic interactions in strict analogy to short-range excluded
  volume interactions. Quite interestingly, the results are consistent
  with the scaling picture of de Gennes et al.~\cite{GPVB_jp_76} in
  the two limits of strong and vanishing screening. Moreover, they are
  supported by recent calculations by BJ and
  others~\cite{BarratJoanny_epl_93,Ha_mm_95} who determined the
  persistence length of the blob chain in a variational procedure and
  found $l_e\sim\kappa^{-1}$.
\end{description}
In addition, there is a number of recent theories which fall into
neither of the two classes outlined
above~\cite{VilgisWilder_ctps_98,WilderVilgis_pre_98,LiverpoolStapper_epl_97,LiverpoolStapper_epje_01,HansenPodgornik_jcp_01}.
While there is a growing consensus among theoreticians that the \OSFKK
result is {\em asymptotically}
correct~\cite{LiWitten_mm_95,Ha_jcp_99,NetzOrland_epjb_99},
experiments~\cite{Tricot_mm_84,Reed_mm_91,Foerster_jpc_92,FoersterSchmidt_aps_95,Beer_mm_97}
as well as computer
simulations~\cite{Micka_pre_96,Ullner_jcp_97,Ullner_mm_02} have
consistently provided evidence for a comparatively weak
$\kappa$-dependence of the electrostatic presistence length.
%a non-perturbative $1/d$-expansion~\cite{HansenPodgornik_jcp_01}
%yielding $l_e\sim\kappa^{-7/6}$, a self-consistent variational
%computation of an effective field
%theory~\cite{VilgisWilder_ctps_98,WilderVilgis_pre_98}, and a
%field-theoretic renormalization group
%analysis~\cite{LiverpoolStapper_epl_97,LiverpoolStapper_epje_01}.

The purpose of the present paper is to shed some new light on this
problem by combining a scaling analysis
with large scale Monte Carlo simulations.  We reexamine the \KMBJ and
the \OSFKK theory in order to extract guidance for the data analysis
and the choice of simulation parameters.  As a result we are able (i)
to rule out the \KMBJ theory, (ii) to provide benchmark results for
analytical solutions of the DHWLC model as well as (iii) for a
comparison to experiments in order to clarify if the systems under
consideration are actually described by the DHWLC model or if
additional effects such as solvent quality or counter-ion condensation
need to be taken into account as well.

The paper is organized as follows: In
section~\ref{sec:choice_of_model} we review the predictions of the
\KMBJ and \OSFKK theories, followed by a discussion in
section~\ref{sec:consequences} of how experiments and simulations
should be set up and analyzed in order to discriminate between the two
scaling pictures. Details of our Monte Carlo simulations can be found
in in section~\ref{sec:method}.  We present our results in
section~\ref{sec:results} and close with a brief discussion.

\section{Scaling theories of intrinsically flexible \DH chains}
\label{sec:choice_of_model}
The DHWLC is characterized by the following set of parameters: $q$,
$l_B$, $\kappa$, $A$, $f$, $l_0$, and $L_{tot}$, where $L_{tot}$
denotes the total chain length and $f\le1$ the fraction of ionized
charged groups which needs to be determined independently for
experimental systems. Setting $f=l_B/A$ is a rough way of accounting
for Manning condensation~\cite{BarratJoanny_acp_96} of counter ions in
cases where $l_B<A$. Correlation functions can be calculated for chain
segments with arbitrary contour length $L<L_{tot}$.  We focus on the
non-OSF limit $l_B,A,l_0\ll \kappa^{-1} \ll \sqrt{\rsq(L_{tot})}$
where the screening length is larger than all microscopic length
scales of the polymer model but smaller than the size of the entire
chain.

For an understanding of the physics, some of these length scales and
parameters are less relevant than others. For example, the actual
distribution of the charges on the chain should be unimportant as long
as $A\ll\kappa^{-1}$.  In our simulations we use discrete charges
spaced by a distance equal to the intrinsic
persistence length, while the scaling arguments assume a continuous
charge distribution. Similarly, in the non-OSF limit with
$l_0\ll\kappa^{-1}$ the physics should not depend on the details of
the WLC crossover from rigid rod to random coil behavior for $L\approx
l_0$. In our simulations we therefore use freely jointed chains (FJC)
whose (Kuhn) bond length $b$ corresponds (up to a henceforth
neglected factor of two) to the persistence length of a WLC. Finally,
it is convenient to consider the limit of infinite total chain length
in order to eliminate the  $L_{tot}$ dependence.
Again the practical limitation to $N_{tot}=L_{tot}/b=4096$ segments
should be unimportant, since our chains always fulfill the condition
$\kappa^{-1} \ll \sqrt{\rsq(L_{tot})}$.

%For our present purposes, we are left with three independent
%dimensionless parameters: the coupling constant $u=q^2 l_B/(A/f)$
%describing the strength of the electrostatic interactions between
%adjacent charges, the screening constant $(\kappa (A/f))^{-1}$ and the
%segment length $N=L/b$ for which we want to calculate correlation
%functions such as end-to-end distances.

The remaining independent parameters (the line charge density $f q/A$,
$l_B$, $\kappa$, $l_0$ and $L$) can be reduced further using the
notion of a ``polyelectrolyte blob'' which was introduced by de Gennes
et al.~\cite{GPVB_jp_76} to describe the crossover from locally
unchanged chain statistics to stretching on long length scales.

Consider first weakly charged flexible polyelectrolytes, where the
electrostatic interactions are irrelevant on the length scales
comparable to the intrinsic persistence length $l_0$.  On larger
length scales an undisturbed WLC with a contour length $L$ has a
spatial extension $\rsq=2 L l_0$.  Neglecting prefactors, the
electrostatic energy of such a chain is given by 
$U_e/k_BT \simeq q^2 (f L/A)^2 l_B/\sqrt{\rsq}$.  
Electrostatic interactions become
relevant for $U_e/k_BT\ge1$ or chain lengths $L$ exceeding

\begin{eqnarray}
l_g & = & l_0^{1/3} \left(\frac {A^2}{f^2 q^2 l_B}\right)^{2/3}
\label{eq:ge}
\end{eqnarray}
and whose spatial extension is given by
\begin{eqnarray}
\xi &=&  l_0^{2/3} \left(\frac {A^2}{f^2 q^2 l_B}\right)^{1/3} .
\label{eq:xe}
\end{eqnarray}
However, this derivation breaks down for strongly interacting systems
where the contour length per blob becomes smaller than the intrinsic
persistence length.  In this case, a similar argument can be made for
a WLC with $L<l_0$ and $\rsq= L^2$ yielding

\begin{equation}
l_g = \xi = \frac {A^2}{q^2 f^2 l_B}.
\label{eq:gexe}
\end{equation}
Both definitions match for $l_g=l_0$, hiding a subtle
crossover~\cite{Ha_jcp_99} behind a crudely renormalized system of
units. Throughout the paper all quantities will be expressed using
these natural units of contour length and spatial distance. On a
scaling level, they become a function of only {\em two} dimensionless
parameters: the reduced chain or segment length

\begin{equation}\label{eq:X}
X  = L/l_g
\end{equation}
and the reduced screening length

\begin{equation}\label{eq:Y}
Y = (\kappa \xi)^{-1} \ .
\end{equation}
%While our extension Eq.~(\ref{eq:gexe}) of the blob scaling 
%to strongly charged polyelectrolytes is plausible,  its
%verification of the validity of this approach  

\begin{figure}[t]
  \begin{center}
    \epsfig{file=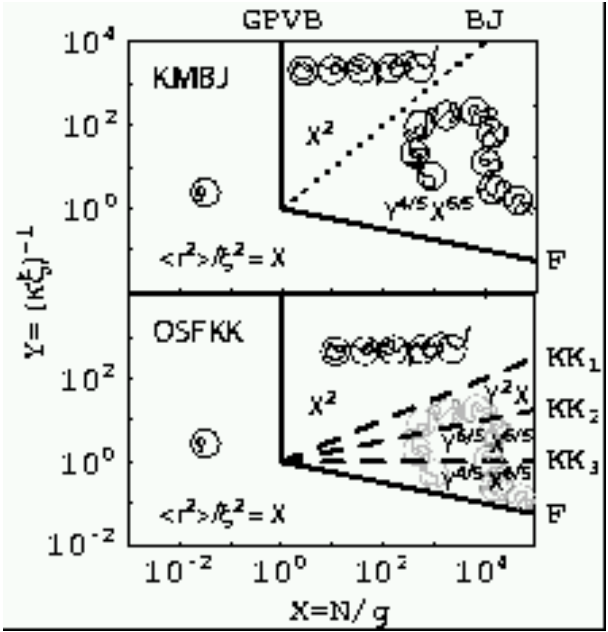,width=8cm}
    \caption{ \label{fig:BJKKMap} 
      Conformations of DHWLC beyond the OSF limit  as a function of
      reduced chain length and reduced screening length.
      Electrostatic interactions are relevant in the upper right part of
      the map which is limited by solid black lines.
%      The solid black lines indicate the chain length beyond which
%      electrostatic interactions become relevant. 
      The dashed and dotted lines correspond to crossover lines in the
      \OSFKK and \KMBJ scaling theories respectively. We have also
      included the scaling predictions Eqs.~(\protect\ref{eq:r2_blob})
      to (\protect\ref{eq:r2_SAW_KK}) for the chain radii.  For
      details see the main text.}
  \end{center}
\end{figure}

In the model under consideration the electrostatic interactions are
purely repulsive.  Therefore the chains are always extended relative
to the neutral case.  For $\kappa l_B\ll 1$ the details of the process
are quite involved and difficult to treat from first principles.  The
approximation schemes used are often based on mechanical analogies
such as stretching, bending and swelling due to short-range excluded
volume interactions and have been reviewed in
Ref.~\cite{BarratJoanny_acp_96,Ullner_handbook_02}.  We have
summarized the various predictions in terms of a schematic map of the
$XY$ parameter space (Figure~\ref{fig:BJKKMap}).  Before we discuss
the controversial parts, we first present those aspects which seem
well understood:

\begin{itemize}
\item The \DH interaction is irrelevant inside the
electrostatic blob, i.e. for weakly charged chains

\begin{equation}\label{eq:r2_blob}
\rsq/\xi^2 \simeq X \ \ \ \ \ \mbox{for $X<1$.}
\end{equation}
For strongly charged chains, this regime does not exist.
\item In the absence of screening, when the monomers interact via an
{\em infinite range} Coulomb potential, the chains are stretched
into a ``blob pole'':

\begin{equation}\label{eq:r2_blobpole}
\rsq/\xi^2 \simeq X^2 \ \ \ \ \ \mbox{for $1<X<\infty$ and $Y\rightarrow\infty$}
\end{equation}
In Figure~\ref{fig:BJKKMap} we have marked the line dividing these
two regimes as ``GPVB'' after the initials of the authors of
Ref.~\cite{GPVB_jp_76} where the notion of the electrostatic blob was
introduced.

\item For sufficiently long chains, the \DH interaction becomes effectively
short--ranged, leading to self-avoiding walk (SAW) behavior

\begin{equation}\label{eq:r2_SAWgeneral}
\rsq/\xi^2 \sim X^{2\nu} \ \ \ \ \ \mbox{for $0<Y<\infty$ and $X\rightarrow\infty$}
\end{equation}
where $\nu \approx 3/5$ is the usual Flory exponent.

\item For strong screening with $q^2 l_B<\kappa^{-1}<A$ the \DH interaction
reduces to an ordinary excluded volume potential with a second virial 
coefficient $v\simeq q^2 l_B \kappa^{-2}$ between charges. 
Using a conventional Flory argument to balance the two-body repulsion
$v (f L/A)^2/R_F^3$ with the entropic elasticity of a Gaussian chain
$R_F^2/(L l_0)$, one obtains:

\begin{equation}\label{eq:r2_SAW_Ysmall}
\rsq/\xi^2 \simeq \left\{\begin{array}{ll}
        X & \mbox{for  $X<Y^{-4}$}\\
        Y^{4/5} X^{6/5} & \mbox{for  $X>Y^{-4}$}\\
     \end{array}\right.
\ \ \mbox{for $Y\ll1$}
\end{equation}
In Figure~\ref{fig:BJKKMap} the corresponding line, 
beyond which the short range excluded volume interaction becomes
relevant, is marked as ``F''.
\end{itemize}

The controversial parts of the phase diagram concern the crossover
from the blob pole to the self-avoiding walk regime. The problem is often
treated in analogy to a simple WLC.
%the thermal undulations of worm-like chains
%with a certain bending rigidity $E$, which are stiff with
%$\rsq = L^2$ on length scales below the persistence length
%$l_p = E/k_BT$ and flexible with $\rsq\simeq L l_p$ for $L\gg l_p$.
With an Onsager virial coefficient $v\simeq l_p^2 d$ between rigid segments
of length $l_p$ and diameter $d$, excluded volume effects become relevant 
beyond a ``Flory length'' 
$l_F=l_p^3/d^2$ leading to a Flory radius of
$R_F^2 \simeq d^{2/5} l_p^{8/5} (L/l_p)^{6/5}$. 
In the case of the blob chain, the diameter of the electrostatically
excluded volume is given by $d\simeq\kappa^{-1}$
\cite{Odijk_jpspp_78,FixmanSkolnick_mm_78}. However, there 
is disagreement with respect to the $\kappa$--dependence 
of the electrostatic persistence length $l_e$. 

\begin{itemize}
\item Variational approaches such as the theory of Barrat and Joanny
  (BJ) often predict $l_e \simeq d \simeq \kappa^{-1}$. As a
  consequence, $l_F = l_p =\kappa^{-1}$ so that there is a direct
  crossover from the stiff blob chain to a SAW regime when the contour
  length $\xi \frac Ng$ of the blob chain reaches the screening length
  $\kappa^{-1}$. Using dimensionless units this corresponds to
  $X= N/g=l_F/\xi = l_p/\xi =1/(\kappa\xi)=Y$ (the dotted line
  in Figure~\ref{fig:BJKKMap} marked ``BJ''). The result 

\begin{equation}\label{eq:r2_SAW_BJ}
\rsq/\xi^2 \simeq Y^{4/5} X^{6/5} \ \ \ \ \ \mbox{for $X>Y$ and $Y\gg1$}
\end{equation}
is identical to Eq.~(\ref{eq:r2_SAW_Ysmall}). On a scaling level,
the predictions of the BJ theory coincide with those of the 
excluded volume theories of Katchalsky~\cite{Katchalsky_ahca_48} and
 Muthukumar~\cite{Muthukumar_jcp_87,Muthukumar_jcp_96,Muthukumar_jcp_01}.
\item
Most theories favour the relation $l_e\simeq \xi^{-1}\kappa^{-2}$ first
obtained by Khokhlov and Khachaturian (KK). KK argued that the OSF
result Eq.~(\ref{eq:OSF}) should also apply to a stretched
chain of blobs with line charge density $f q l_g/(A\xi)$
so that $l_e/\xi = 1/(\kappa \xi)^2=Y^2$.
Since $l_e\gg d$, the resulting phase diagram is considerably
more complicated. The blob chain starts to bend for
reduced segment lengths $X$ exceeding $l_p/\xi = l_e/\xi = Y^2$
(the ``$\mathrm{KK}_1$'' line in Figure~\ref{fig:BJKKMap}),
while excluded volume effects become
relevant beyond  
$l_F/\xi= 1/(\kappa \xi)^4 =Y^4$ (``$\mathrm{KK}_2$'').

\begin{equation}\label{eq:r2_SAW_KK}
\rsq/\xi^2 \simeq \left\{\begin{array}{ll}
        Y^2 X & \mbox{for  $Y^{2}<X<Y^{4}$}\\
        Y^{6/5} X^{6/5} & \mbox{for  $X>Y^{4}$}\\
     \end{array}\right.
\ \ \mbox{for $Y\gg1$}
\end{equation}
Finally, and in contrast to the \KMBJ theory, the \OSFKK approach
implies another crossover (``$\mathrm{KK}_3$'') within the SAW regime at $Y=1$
from Eq.~(\ref{eq:r2_SAW_KK}) to  Eq.~(\ref{eq:r2_SAW_Ysmall}).
\end{itemize}

\section{Implications for Data Production and Analysis}
\label{sec:consequences}
In general, scaling theories make two kinds of predictions: (i) about
the existence of characteristic length scales or crossover lines in
conformation space and (ii) about the {\em asymptotic} behavior of
observables in the areas between these crossover lines.  In principle,
attempts at refutation can aim at either type of prediction.
However, in the present case the identification of asymptotic
exponents turns out to be particularly difficult.  Apart from
numerical prefactors and logarithmic corrections~\footnote{For
  example, in the absence of screening the long-range Coulomb
  interactions along the blob pole create a tension which grows
  logarithmically with the chain length $N$.  As a consequence, the
  blob diameter is reduced and the chains grow with $R_g \sim N
  \log^{1/3}(N)$.  In particular, a $N$ monomer segment of a longer
  chain will always be more extended than a $N$
  monomer chain whereby the deformation is strongest for segments
  located near the center of the longer
  polymer~\cite{GPVB_jp_76,CastelnovoSensJoanny_epje_00}.  In the
  presence of screening, this effect leads on the one hand to an
  increase of the contour length of the blob chain.  On the other
  hand, the correspondingly reduced line charge density results in a
  reduction of the \OSFKK persistence length.  Similarly, but neglecting
  the stretching, there are logarithmic corrections to the
  electrostatically excluded volume around the blob
  chain~\cite{FixmanSkolnick_mm_78}.  Moreover, although fairly
  robust, the Flory argument used to estimate the excluded volume
  effects is far from exact.  In addition to the aforementioned
  crossovers, a complete theory will have to account for all of these
  effects.} one is faced with four problems:
\begin{itemize}
\item Although the various regimes predicted by the \KMBJ and the \OSFKK
  theory are characterized by different combinations
  of powers of $X$ and $Y$, the exponents are often similar and the
  {\em absolute} differences between the predicted chain extensions
  relatively small.
\item At least on a scaling level all crossover lines meet at $X=Y=1$
  for chain and screening length of the order of the diameter of the
  polyelectrolyte blob. In the case of the \OSFKK theory some of the
  predicted regimes are extremely narrow in the sense that chain
  lengths of $X=10^4$ blobs are required for a width of one order of
  magnitude in $Y$-direction. This validity range would seem to be the
  absolute minimum for identifying power law behavior.
\item The crossover lines are neither parallel to each other nor to
  the ``natural'' $X$ and $Y$ directions of variations of chain 
  and screening length respectively.
\item In particular, results will be influenced by the finite total
  length $L_{tot}$ of polyelectrolyte chains studied in experiments or
  simulations. The importance of these effects varies with the ratio
  of the screening length $Y$ and the contour length $X$ of the blob
  chain. As a consequence, they risk to mask the asymptotic
  $Y$-dependence of observables such as the electrostatic persistence
  length, if they are evaluated for chains with {\em fixed}  $L_{tot}$. 
\end{itemize}
Quite obviously, the discrimination between the two scaling pictures
requires the investigation of chains whose {\em effective} length $X$ is as
large as possible.  In addition one should rely on those observables and
data representations which are {\em most} sensitive to the differences
between the theories and {\em least} sensitive to the omitted constants,
corrections and crossovers.  In the following we discuss the analysis
of data for internal distances and for the tangent correlation function. 
In particular, we will argue that it is relatively easy to discard the \KMBJ
picture using simple scaling plots, while the verification of
some of the predictions by the \OSFKK
theory requires astronomical chain lengths.

\begin{figure}[t]
  \begin{center}
    \epsfig{file=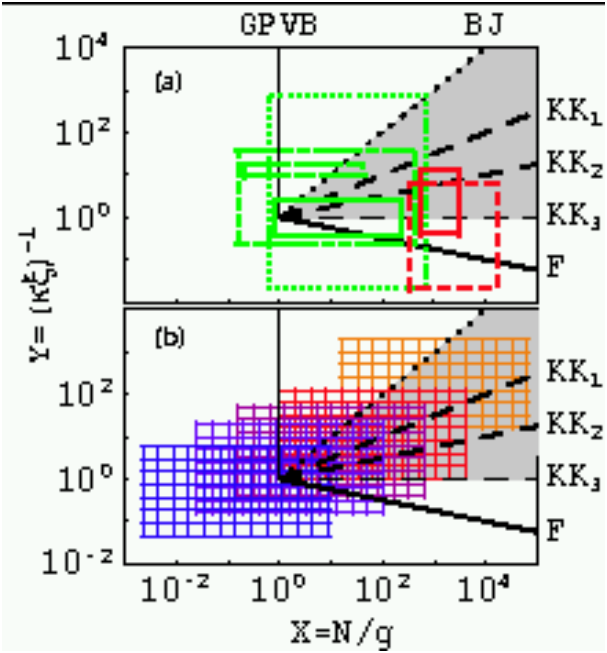,width=8cm}
    \caption{ \label{fig:SimExpXYMap} 
      Areas of the conformation diagram Fig.~\protect\ref{fig:BJKKMap}
      for which there are experimental and simulation data available.
      In (a) the green lines indicate the parameter ranges
      investigated in previous numerical studies: Barrat and
      Boyer~\protect\cite{BarratBoyer_jp2_93} (--- ---),
      Seidel~\protect\cite{Seidel_bunsen_96} (------), J\"onsson et
      al.~\protect\cite{Jonsson_jpc_95,Ullner_jcp_97}
      ($\cdots\cdots$), Micka and Kremer~\protect\cite{Micka_pre_96}
      (-- -- --). Experiments (shown in red) have access to longer
      chains, but the reduced screening length $Y$ are typically
      smaller than ten: Reed et al.~\protect\cite{Reed_mm_91} (-- --
      --), Beer et al.~\protect\cite{Beer_mm_97} (------).  The
      colored grids in (b) denote the parameter ranges covered by
      different sets of our MC simulations. Note that the predictions
      of the \KMBJ and the \OSFKK theory differ most strongly along the
      $\mathrm{KK}_1$ line and are identical outside of the the gray
      shaded area.  }
  \end{center}
\end{figure}

Consider first the scaling predictions Eqs.~(\protect\ref{eq:r2_blob})
to (\protect\ref{eq:r2_SAW_KK}) for the mean square internal distances
at reduced screening lengths $Y>1$.  Describing the GPVB crossover to
the blob pole within the chain-under-tension
model~\cite{GPVB_jp_76,BarratBoyer_jp2_93}, Eqs.~(\ref{eq:r2_blob}) and
(\ref{eq:r2_blobpole}) can be combined as $\rsq/\xi^2=X+X^2$.  Taking
this into account, the \KMBJ theory suggests that {\em all} data points
should collapse when plotted in the following manner as a function of
the \KMBJ persistence length:

\begin{equation}\label{eq:r2BJScaling}
\frac{\langle r^2(Z=X/Y)\rangle/\xi^2-X}{X^2}
\simeq 
\left\{\begin{array}{ll}
        1 & \mbox{for  $Z<1$}\\
        Z^{-4/5} & \mbox{for  $Z>1$}\\
     \end{array}\right.
\end{equation}
In contrast, the \OSFKK theory predicts data collapse, if the segment lengths
are rescaled with the \OSFKK persistence length $Y^2$, and {\em a breakdown
of scaling} for segment lengths approaching
the Flory length $Y^4$:

\begin{equation}\label{eq:r2KKScaling}
\frac{\langle r^2(Z=X/Y^2)\rangle/\xi^2-X}{X^2}
\simeq 
\left\{\begin{array}{ll}
        1 & \mbox{for  $Z<1$}\\
        Z^{-1} & \mbox{for  $1<Z<Y^2$}\\
        Y^{-2/5}Z^{-4/5} & \mbox{for  $Y^2<Z$}\\
     \end{array}\right.
\end{equation}
The predictions of the two scaling theories differ
most strongly for chain radii along the
$Y=X^{1/2}$ $\mathrm{KK}_1$ line for segment lengths equal to the \OSFKK
persistence length.  In the \KMBJ theory, this line is already deep in
the SAW regime. Eqs.~(\ref{eq:r2_SAW_BJ}) and (\ref{eq:r2_SAW_KK}) imply

\begin{equation}\label{eq:r2_XeqY2}
\frac {\langle r^2(X,Y=X^{1/2})\rangle}{X\xi^2} \simeq 
\left\{\begin{array}{ll}
         X^{3/5} & \mbox{(\KMBJ)}\\
        X & \mbox{(\OSFKK)}\\
     \end{array}\right.
\end{equation}
Thus the ratio
$\rsq_{\OSFKK}/\rsq_{\KMBJ}=X^{2/5}$ is fairly small and
$X=10^{5/2}$ blobs are required for this ratio to become of order
10.  
For comparision, both theories predict full extension along the \KMBJ line

\begin{equation}\label{eq:r2_XeqY}
\frac {\langle r^2(X,Y=X)\rangle}{X\xi^2} \simeq 
\left\{\begin{array}{ll}
        X & \mbox{(\KMBJ)}\\
        X & \mbox{(\OSFKK)}\\
     \end{array}\right.
\end{equation}
and differ only by a factor of $\rsq_{\OSFKK}/\rsq_{\KMBJ}=X^{1/10}$ 
along the $\mathrm{KK}_2$ line 

\begin{equation}\label{eq:r2_XeqY4}
\frac {\langle r^2(X,Y=X^{1/4})\rangle}{X\xi^2} \simeq 
\left\{\begin{array}{ll}
        X^{2/5} & \mbox{(\KMBJ)}\\
        X^{1/2} & \mbox{(\OSFKK)}\\
     \end{array}\right.
\end{equation}

While segment lengths of the order of $X=10^3$ blobs are thus
sufficient to discriminate between the \KMBJ and the \OSFKK proposals
for the electrostatic persistence length, the requirements
for resolving the additional crossovers predicted by the \OSFKK 
theory are much higher.
%Compared to a data analysis which traces properties {\em along} the
%predicted crossover lines,  crossover scaling {\em across}
%these lines will run into even more severe problems.
Consider again the $\mathrm{KK}_2$ line where excluded volume
effects are expected to become relevant for the undulating blob chain.
Eq.~(\ref{eq:r2_SAW_KK}) can be rewritten in the form

\begin{equation}\label{eq:r2_SAW_KK_Z}
\frac{\langle r^2(Z=Y^{-4/5}X^{1/5})\rangle}{Y^2 X\xi^2} \simeq 
\left\{\begin{array}{ll}
        1 & \mbox{for  $Z<1$}\\
        Z & \mbox{for  $Z\ge 1$}\\
     \end{array}\right.
\end{equation}
where $Z$ is the variable measuring the effective
distance from the crossover line at $Z=1$. In order to identify the
asymptotic behavior one needs at least data covering the interval
$Z\in [0.1,10]$. Since the validity range of
Eq.~(\ref{eq:r2_SAW_KK_Z}) is limited by the $\mathrm{KK}_1$ and
$\mathrm{KK}_3$ lines (so that $1<Y<X^{1/2}$ or
$X^{1/5}>Z>X^{-1/5}$), this implies a minimum segment length of $X=10^5$
blobs for establishing the $\mathrm{KK}_2$ crossover.  Similarly the
$\mathrm{KK}_3$ crossover between Eqs.~(\ref{eq:r2_SAW_Ysmall})
and (\ref{eq:r2_SAW_KK}) at $Y=1$ becomes relevant for chains of
$X=10^{10}$ blobs! 

The mean-square internal distances and the tangent correlation
function (TCF) obey a Green-Kubo like relation:

\begin{equation}\label{eq:BACF_GreenKubo}
\langle {\vec b}_N \cdot {\vec b}_0 \rangle =\frac12
\frac {d^2}{d\, N^2} \langle r^2(N) \rangle =\frac12
\frac {\xi^2}{g^2} \frac {d^2}{d\, X^2} \frac {\langle r^2(X) \rangle}{\xi^2}
\end{equation}
For a WLC the TCF is simply given by

\begin{equation}\label{eq:bacf}
\langle {\vec b}(s) \cdot {\vec b}(0) \rangle = b^2 \exp(-s/l_p)
\end{equation}
so that the persistence length can be read off directly from a
semi-logarithmic plot.  Numerical studies of
polyelectrolytes~\cite{Micka_pre_96,Ullner_jcp_97,Ullner_mm_02} have
therefore often focused on this quantity in spite of two intrinsic
problems: (i) the TCF is considerably more difficult to measure with
the same relative precision than internal distances and (ii) the TCF
is particularly sensitive to finite chain length effects (a
characteristic sign is the {\em faster} than exponential decay of the
TCF on length scale approaching the chain length). In contrast to the
case of ordinary SAWs~\cite{Schaefer_jp_99}, nothing is known about
the functional form of the corrections. In the following discussion we
will focus on a third aspect: the sensitivity of the TCF to the
neighborhood of crossover lines.

On a scaling level, the behavior of the TCF can be obtained by 
applying Eq.~(\ref{eq:BACF_GreenKubo}) to 
Eqs.~(\protect\ref{eq:r2_blob}) to (\protect\ref{eq:r2_SAW_KK}).
For $Y>1$ the \KMBJ theory predicts

\begin{equation}\label{eq:bacfBJ}
\frac {g^2}{\xi^2} \langle {\vec b}(Z=X/Y) \cdot {\vec b}(0) \rangle \simeq 
\left\{\begin{array}{ll}
        1 & \mbox{for  $Z<1$}\\
        Z^{-4/5} & \mbox{for  $Z>1$}\\
     \end{array}\right.
\end{equation}
Within the \OSFKK theory, simple predictions can only be made 
for segment lengths below the persistence length and beyond
the Flory length:

\begin{equation}\label{eq:bacfKK}
\frac {g^2}{\xi^2} \langle {\vec b(Z=X/Y^2)} \cdot {\vec b}(0) \rangle \simeq 
\left\{\begin{array}{ll}
        1 & \mbox{for  $Z<1$}\\
        Y^{-2/5} Z^{-4/5} & \mbox{for  $Y^2<Z$}\\
     \end{array}\right.
\end{equation}
However, since both theories are based on the analogy to a mechanical
WLC, they are often associated with the much more detailed prediction

\begin{eqnarray}\label{eq:bacfWLC}
\lefteqn{\frac {g^2}{\xi^2} \langle {\vec b}(X) \cdot {\vec b}(0) \rangle \simeq }\\
&&
\left\{\begin{array}{ll}
        \exp(-X/Y)   & \mbox{for  $X<Y$\ \ \ (\KMBJ)}\\
        \exp(-X/Y^2) & \mbox{for  $X<Y^2$\ \ \ (\OSFKK)}\\
     \end{array}\right.
\nonumber
\end{eqnarray}
for the functional form of the decay of the tangent correlations.
Measuring this quantity for DHWLC therefore seems to be the most
direct way of justifying or refuting this analogy and its
exploitation.  In particular, numerical
work~\cite{Micka_pre_96,Ullner_jcp_97,Ullner_mm_02} has concentrated on (i)
establishing the existence of a single exponential decay of the
TCF over a certain range of length scales and
(ii) extracting the $\kappa$-dependence of the measured decay length.
In the following we will reexamine this approach by taking a closer
look at Eqs.~(\ref{eq:bacfBJ}) and (\ref{eq:bacfKK}), since they
contain additional crossovers neglected in Eq.~(\ref{eq:bacfWLC}).

The situation should be uncritical for the GPVB crossover where the
chain-under-tension models~\cite{GPVB_jp_76,BarratBoyer_jp2_93}
suggests that Eq.~(\ref{eq:bacfWLC}) remains valid for $X<1$. In
contrast, nothing is known in detail about the way the TCF crosses
over to the slow power law decay characteristic for the SAW behavior on
large length scales.  However, matching Eq.~(\ref{eq:bacfWLC}) (which
only accounts for the local bending rigidity) with the asymptotic
behavior in Eqs.~(\ref{eq:bacfBJ}) and (\ref{eq:bacfKK}) shows that
the tangent-correlation function is much more sensitive to excluded
volume effects than the chain radii.  This is most obvious for the \OSFKK
theory where the two limits match close to the \OSFKK persistence length
$X=Y^2$ instead of the Flory length $X=Y^4$.  While the scaling of the
TCF with the \OSFKK persistence length should start to break down around
$X/Y^2\approx 1$, one can nevertheless expect Eq.~(\ref{eq:bacfWLC})
to hold up to this point.  In the case of the \KMBJ theory the situation
is quite different, since the persistence length and the Flory length
coincide.  As a consequence, Eq.~(\ref{eq:bacfWLC}) effectively breaks
down as soon as the tangent-correlation function starts to deviate
from one. On the other hand, in the absence of other relevant length
scales the TCF should scale with the \KMBJ persistence length for {\em
  arbitrary} segment length!

In our opinion, these arguments shed some doubts on attempts to
identify the electrostatic presistence length which are based
too closely on Eq.~(\ref{eq:bacfWLC}). Scaling plots testing
Eqs.~(\ref{eq:bacfBJ}) and (\ref{eq:bacfKK}) may offer a simpler
{\em and} safer alternative.

\section{Simulation Model, Method and Parameters}
\label{sec:method}
\begin{figure*}[t]
  \begin{center}
    \epsfig{file=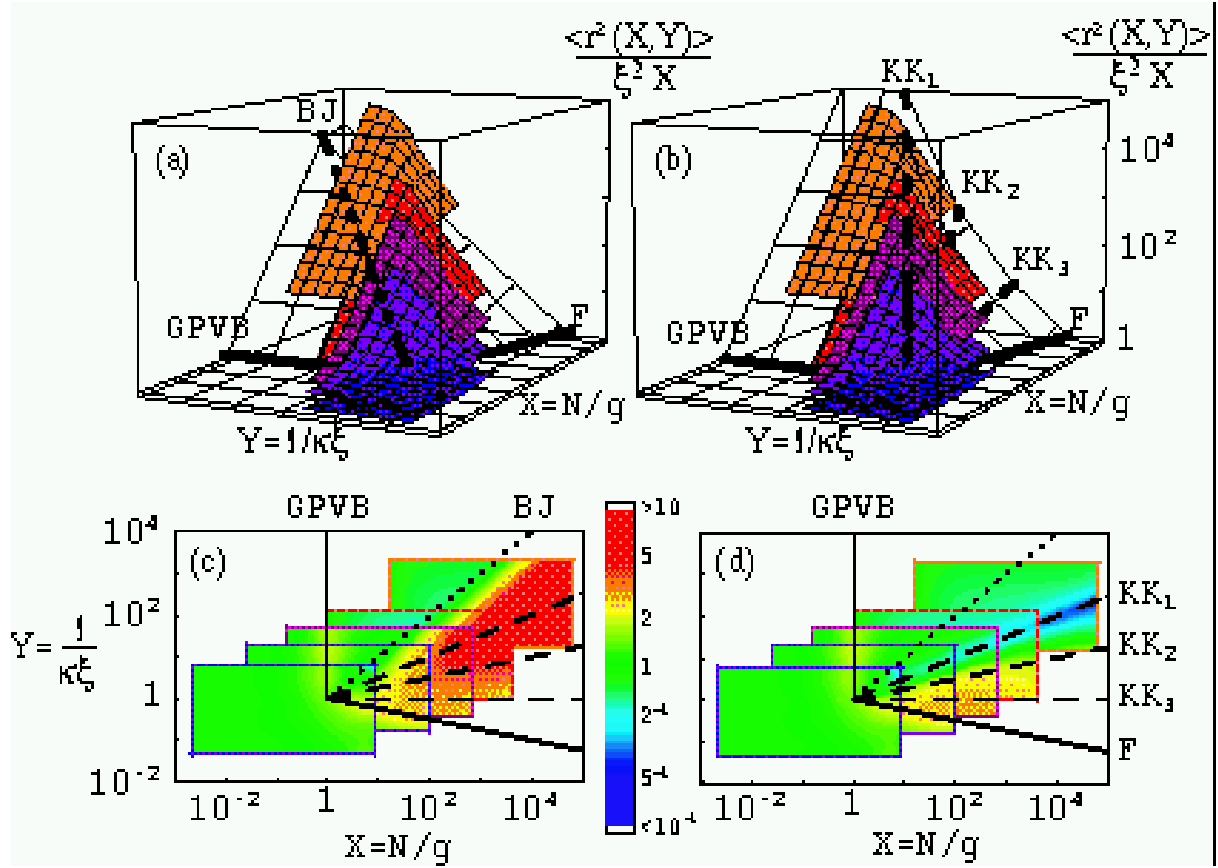,width=16cm}
    \caption{ \label{fig:r2OverXPlot} 
      Comparison of measured internal distances $\langle
      r^2(X,Y)\rangle/\xi^2$ to the predictions of the \KMBJ ((a) and
      (c)) and \OSFKK ((b) and (d)) scaling theories. In the top row we
      show log-log-log representations where all distances are
      normalized to the undisturbed random walk. The colored areas
      were generated by interpolation between the results of all
      simulations for a given coupling strength.  The supporting grids
      and the crossover lines show the two sets of scaling predictions
      Eqs.~(\protect\ref{eq:r2_blob}) to (\protect\ref{eq:r2_SAW_KK})
      as an extensions of Fig.~(\protect\ref{fig:BJKKMap}) to three
      dimensions.  In the plots of the bottom row the colors indicate
      the ratios $\langle r^2(X,Y)\rangle/\rsq_{\OSFKK}$ and $\langle
      r^2(X,Y)\rangle/\rsq_{\KMBJ}$ respectively.  }
  \end{center}
\end{figure*}
%\cleardoublepage

As already mentioned in section~\ref{sec:choice_of_model},
we model the polymers as freely jointed chains (FJC) with unit
charges $q=1$ at each joint.
Lengths were measured in units of the bond length $b$.
We varied the the Bjerrum length
$l_B=4^2,1,1/4^2,1/16^2,1/100^2b$
and the screening length
$\kappa^{-1}= 1,2,4,8,16,32,64,128\ b$.
%All monomers in the simulation were equally charged
%with $q=4,1,1/4,1/16,1/100$, which can be interpreted
%as distances  $A/b=1/4,1,4,16,100$ between unit charges $q_A=1$.
As in our previous study on
poly\-ampholytes~\cite{Yamakov_prl_00}, the chains have a length of 
up to $N=4096$ monomers.  
\begin{itemize}
\item Since we study the conformations of isolated chains, we employ
  the efficient technique of pivot rotations due to Sokal et
  al.\cite{MadrasSokal_jstatphys_88,Jonsson_jpc_95}. We use two types
  of pivot moves: Either we rotate the part between the free end of
  the chain and a randomly selected monomer around an axis, defined by
  the bond between this monomer and its nearest neighbor; or we rotate
  a segment between two randomly selected monomers around an axis
  joining them.  The latter provides better efficiency in the case of a
  stretched chain with large excess charge. One MC step consists of
  $N$ attempted rotations at random positions along the chain. Chain
  conformations are stored at intervals of 8-32 MC steps.  For each
  parameter set we simulate 8 independent Markov chains in parallel.
  We typically store $8\times60$ conformations representing a total of
  $1.5\times 10^7$ attempted rotations for our longest chains.
\item Instead of the slower procedure of Stellman and
  Gans\cite{Stellman_mm_72} which corrects the accumulating numeric error in
  off-lattice implementations of the pivot algorithm, we regularly
  reconstruct the chains with the correct bond length.
\item For calculating the long-range electrostatic interactions we
  use a direct summation whereby the energy of the system is obtained
  by direct counting of all the pair energies of the beads. This
  method is still efficient for macromolecules of up to few thousand
  monomers. The DH potential is tabulated in two
  arrays for short and long distances respectively.
\item For better efficiency starting configurations of the chains are
  created by means of the configurational biased\cite{FrenkelSmit_96}
  MC method although one should keep in mind that due to the long
  range interactions the first part of the newly grown chain does not
  experience the cumulative field of the rest of the chain and a
  number of rotational moves are still needed before the chain is well
  equilibrated. Measurements are performed and conformations stored
  only after the chain end-to-end distances are well equilibrated. 
\item Since statistics is gathered both with respect to chain
  conformations as well as to different Bjerrum length $l_B$ and
  screening length $\kappa^{-1}$, we use a simple parallelization
  where different processors of a CrayT3E supercomputer perform
  independent simulation of single chains.  The total CPU time used
  for this project is of the order of $1.5 10^5$ single processor hours.
\end{itemize}
The simulation parameters translate into 
our blob units as

\begin{eqnarray}
g & = &\left(\frac b{l_B}\right)^{2/3} \ \ \ \ \ \ (g>1)
\label{eq:ge_large}\\
\xi &=& b \left(\frac b{l_B}\right)^{1/3}\ .
\label{eq:xe_large}
\end{eqnarray}
and 

\begin{eqnarray}
g & = &\frac b{l_B} \ \ \ \ \ \ (g\le1)
\label{eq:ge_small}\\
\xi &=& b \frac b{l_B}\ .
\label{eq:xe_small}
\end{eqnarray}
where we now use the number $g$ of monomers per blob instead
of the corresponding contour length $l_g=b g$.

Fig. (\ref{fig:SimExpXYMap}) shows where our own data are located within the
$XY$-conformation space. 
%We actually keep the grid representation of our
%data (typically spaced by powers of 2 in $X$- and $Y$-direction)
%throughout much of the data analysis, since the grid lines represent
%two typical experimental situations: variation of chain lengths under
%constant conditions and variation of the screening length for a given
%polymer sample.
The effective chain and screening lengths studied cover a range of
seven and five orders of magnitude respectively. The reduced mean
square internal distances vary over ten orders of magnitude. Along the
$\mathrm{KK}_1$ line our data extend {\em on a logarithmic scale}
about a factor of two further into the asymptotic regime than previous
studies.  While this allows us to discriminate between the \KMBJ and the
\OSFKK predictions for the electrostatic presistence length, our chains
are still too short to resolve the different RW and SAW regimes
predicted by the \OSFKK theory.

Note, that only by studying strongly stretched chains
we are able to  push the {\em effective} chain
length $X$ close to $10^5$ and that our unified description
of strongly and weakly charged flexible polyelectrolytes
needs to be confirmed by the data analysis. To facilitate the
comparison, all figures make use of the same color code to indicate
data obtained for a particular coupling strength $l_B/b$ ranging from
blue for $g=10000^{2/3}\approx470$ over different shades of violet
for $g=256^{2/3}\approx40$ and $g=16^{2/3}\approx6.4$ to red for
$g=1$ and orange for $g=1/16$.  The first three systems can safely
be regarded as Gaussian chains, while the last two are at and beyond
the crossover to the strong stretching regime.

\section{Results}
\label{sec:results}

\begin{figure}[t]
  \begin{center}
    \epsfig{file=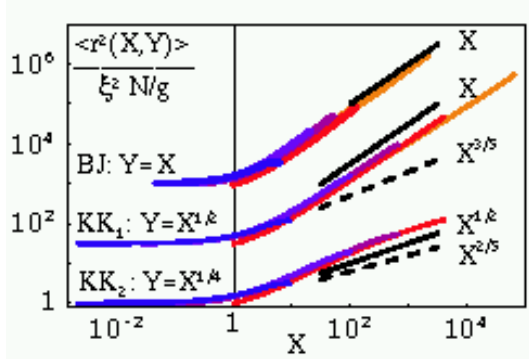,width=7cm}
    \caption{ \label{fig:r2XeqY2} 
      Extension of chain segments along the BJ, $\mathrm{KK}_1$, and
      $\mathrm{KK}_2$ crossover lines in comparison to the predictions
      of the \OSFKK (solid line) and the \KMBJ (dashed line) scaling
      theories (see Eqs.~(\protect\ref{eq:r2_XeqY}),
      (\protect\ref{eq:r2_XeqY2}), and (\protect\ref{eq:r2_XeqY4});
      note that we have not used additional prefactors for this
      comparison).  Results for different coupling constants are
      shifted by factors of $\sqrt{1000}$. }
  \end{center}
\end{figure}

\begin{figure}
  \begin{center}
    \epsfig{file=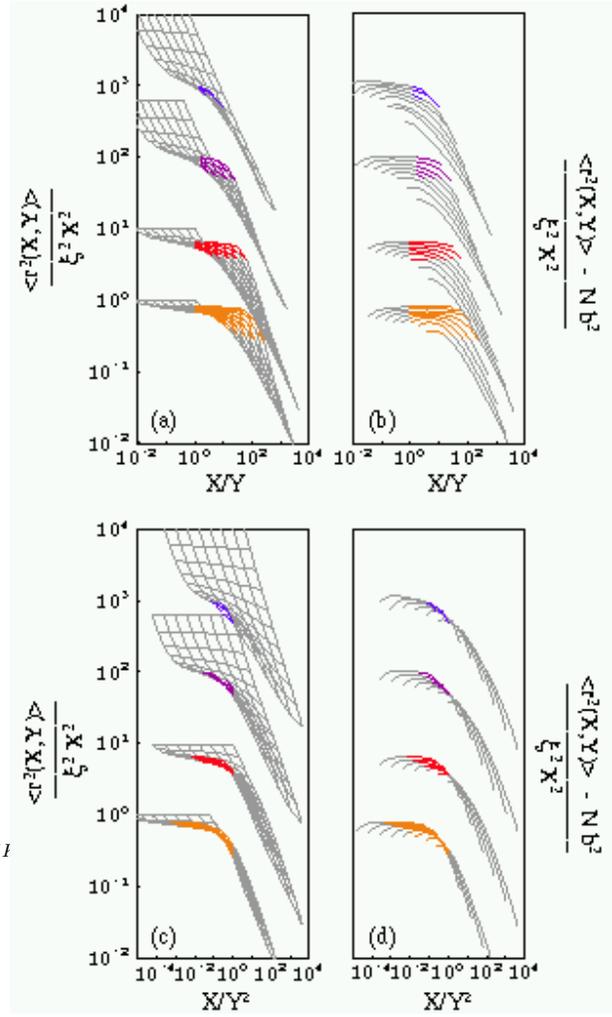,width=8cm}
    \caption{ \label{fig:LpCrossover} 
      Crossover scaling for internal distances versus segment length.
      In the top (bottom) row segment lengths $X$ are normalized to
      the \KMBJ (\OSFKK) persistence length $Y$ ($Y^2$) respectively.
      Figures (a) and (c) on the left-hand side show internal
      distances $\langle r^2(X,Y)\rangle$ normalized to the
      mean-square extension $\xi^2 X^2$ of the blob pole. The grid is
      the same as in Fig.~\ref{fig:SimExpXYMap}(b) and shows $const-X$ and
      $const-Y$ lines.  Figures (b) and (d) on the right-hand side are
      inspired by the chain-under-tension model for Gaussian chains
      Eqs.~(\protect\ref{eq:r2BJScaling}) and
      (\protect\ref{eq:r2KKScaling}). Only $const-X$ lines are shown.
      Data points falling into the range $1<Y<X<Y^2$ are marked using
      the color code indicating the coupling strength.  Results for
      different coupling constants are shifted by factors of ten. }
  \end{center}
\end{figure}

In the data analysis we mainly concentrate on identifying the scaling
behavior: (i) Do different data sets overlap when rescaled according
to our extension Eq.~(\ref{eq:gexe}) of
the definition of the electrostatic blob?  (ii) How do the results of
our simulations compare to the predictions of the \KMBJ and \OSFKK scaling
theories?  In terms of observables we start by presenting data for
internal distances averaged along our chains of total length $N=4096$.
In the second part, we discuss results for the tangent correlation
function. While the TCF was also averaged along the chain, we only
take into account distances up to half the chain length in order
to reduce finite-chain length effects~\cite{Schaefer_jp_99}.  Except
for the most weakly charged system, the chain extensions are much
larger than the screening length, so that we do not expect finite
chain length effects to be very important.  At the end, we briefly
present results for shorter chain lengths.

Figs.~\ref{fig:r2OverXPlot}(a) and (b) are
three-dimensional log-log-log plots giving an overview of all data.
We show reduced internal distances $\langle
r^2(X,Y)\rangle/((N/g)\xi^2)$ normalized to the size of 
the undisturbed random walk as a function of the reduced chain and
screening lengths Eqs.~(\ref{eq:X}) and (\ref{eq:Y}) respectively.
Results for different coupling constants are combined into colored
surfaces, while the supporting grids show the two sets of scaling
predictions Eqs.~(\ref{eq:r2_blob}) to (\ref{eq:r2_SAW_KK}) as an
extensions of Fig.~\ref{fig:BJKKMap} to three dimensions.  

The complementary 
Figs.~\ref{fig:r2OverXPlot}(c) and (d)
show the ratios $\langle r^2(X,Y)\rangle/\langle r^2(X,Y)\rangle_{\KMBJ}$
and $\langle r^2(X,Y)\rangle/\langle r^2(X,Y)\rangle_{\OSFKK}$ of the 
interpolated simulation results to the scaling predictions in
a color coding where green, red and blue indicate agreement, 
under- and overestimation by a factor of three or more respectively.
The advantage of this representation is the localization of
deviations in our schematic map of the parameter space.

Qualitatively, the interpretation of Fig.~\ref{fig:r2OverXPlot}
seems clear. There are neither indications for a failure
of the blob scaling nor for significant deviations from the predictions
of the \OSFKK theory. (We emphasize again that we have neglected all numerical
prefactors and that Eqs.~(\ref{eq:r2_blob}) to (\ref{eq:r2_SAW_KK})
treat crossovers in the crudest manner).
In particular, there is no evidence that
the chains start to bend on length scales comparable to the
screening length as predicted by \KMBJ. In the relevant part of
conformation space, the \KMBJ theory systematically underestimates the
observed chain extensions.
%This conclusion is supported by all of the following more
%detailed comparisons.

Nevertheless, Fig.~\ref{fig:r2OverXPlot} could be misleading, since
the rejection of the \KMBJ theory is mainly based on data falling into
the strong-stretching regime, while the theory is meant to apply to
weakly stretched Gaussian chains. Thus so far our conclusions rest on
the assumption that the extension Eq.~(\ref{eq:gexe}) of the blob
scaling to strongly charged chains can be used to {\em extrapolate}
the behavior of weakly charged systems to segment lengths inaccessible
by simulation. How well this assumption is fulfilled is hard to judge
from Fig.~\ref{fig:r2OverXPlot}.  Definite conclusions require a more
detailed analysis.

Fig.~\ref{fig:r2XeqY2} presents chain radii measured along the BJ,
$\mathrm{KK}_1$ and $\mathrm{KK}_2$ crossover lines.  The first point
to note is that in all three cases we observe almost perfect scaling
of data obtained for different coupling constants.
Eqs.~(\ref{eq:r2_XeqY2}) to (\ref{eq:r2_XeqY4}) predict simple
crossovers at the blob size around $X=1$.  In agreement with both
scaling pictures, we observe stretched blob chains along the BJ line.
The most important set of data are the radii measured along the
$\mathrm{KK}_1$ line for chains with a contour length $X=Y^2$ equal to
the \OSFKK persistence length. In agreement with
the \OSFKK theory we find a simple crossover around $X=1$ to stretched
blob chains.  Contrary to the predictions of the \KMBJ theory the radii
are essentially identical to those observed along the BJ line and do
not show SAW behavior.  In particular, the asymptotic slope predicted
by the \OSFKK theory is already observable for weakly charged chains to
which the \KMBJ theory can be applied directly.  The last set of data is
taken along the $\mathrm{KK}_2$ line which marks the onset of excluded
volume effects in the \OSFKK theory.  Here our results are consistent with
the predictions of both theories. This observation is in agreement
with the estimate of a minimum segment length of $X=10^{10}$ blobs for the
difference to become relevant (see Eq.~(\ref{eq:r2_XeqY4})).
%We show these results in order to illustrate that the method of
%plotting data along crossover lines derived from a scaling theory has
%some intrinsic limits due to the fact that the location of the
%crossovers is only known up to a prefactor.

\begin{figure}[t]
  \begin{center}
    \epsfig{file=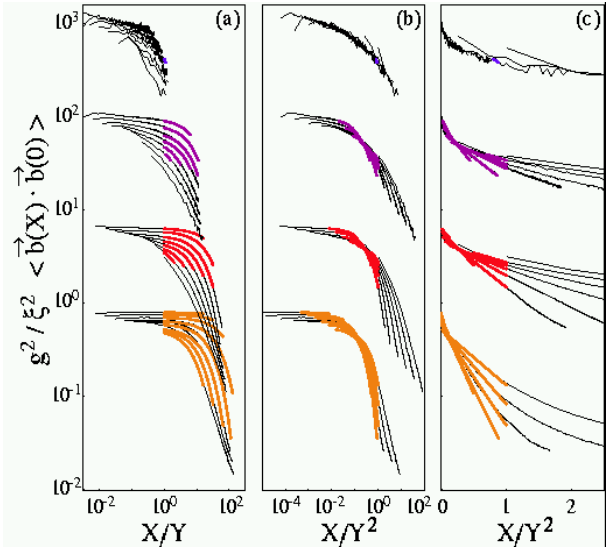,width=8cm}
    \caption{ \label{fig:BACF} 
      Scaled tangent correlation functions. Results for different
      coupling constants are shifted by factors of ten. The colored
      lines mark the results of our fits to a simple exponential decay
      in the range $Y<X<Y^2$. (a) log-log plot using \KMBJ scaling, (b)
      log-log plot using \OSFKK scaling, (c) semi-log plot using \OSFKK scaling. }
  \end{center}
\end{figure}
The screening length dependence of the effective bending rigidity of
the blob chain can also be determined from crossover scaling of
internal distances normalized to the size of the stretched blob pole
(Fig.~\ref{fig:LpCrossover}). The disadvantage of plotting $\langle
r^2(X,Y)\rangle/(\xi^2 X^2)$ directly (Figs.~\ref{fig:LpCrossover} (a)
and (c)) is the occurence of a $1/X$ divergence of results for segment
lengths smaller than the blob size.  Correcting for this in the manner
suggested by the chain-under-tension model
(Eqs.~(\ref{eq:r2BJScaling}) and (\ref{eq:r2KKScaling})) as in
Figs.~\ref{fig:LpCrossover} (b) and (d) largely eliminates effects due
to the GPVB crossover, but introduces some artifacts for small $N$
where $\langle r^2(N=1)\rangle -b^2 \equiv 0$ for a FJC. In agreement
with our previous results and independently of the coupling constant
we observe extremely poor scaling when the data are plotted as a
function of the ratio $X/Y$ of chain length over \KMBJ persistence
length. In constrast, the data superimpose considerably better, if the
\OSFKK scaling is used as in the corresponding
Figs.~\ref{fig:LpCrossover}(c) and (d).  In particular,
Fig.~\ref{fig:LpCrossover} eliminates the possibility of an
electrostatic persistence length scaling like $\kappa^{-1}$ but with
an unusually large prefactor.

\begin{figure}
  \begin{center}
    \epsfig{file=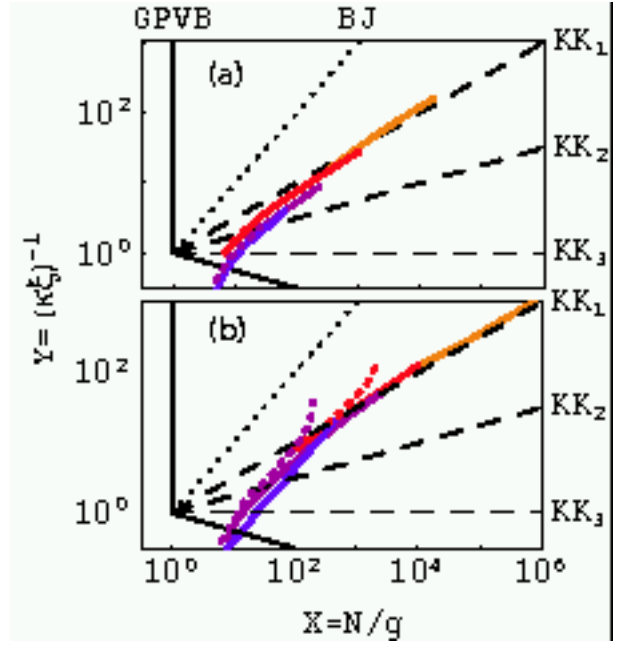,width=8cm}
    \caption{ \label{fig:FittedLp} 
      Location of apparent electrostatic persistence lengths
      $l_{e,app}$ in our schematic map of the $XY$ parameter space. We
      show results for the reduced crossover distance
      $X_{cd}=l_{cd}/\xi$ (a) and the reduced orientational
      correlation length $X_{oc}=l_{oc}\xi$ (b) for chains of
      length $N_{tot}=4096$ (solid lines) and $N_{tot}=256$ (dashed
      lines). In the first case, we compare $\rsq(X,Y)$ to the size of
      the stretched blob pole Eq.~\protect\ref{eq:r2_blobpole} and
      define $X_{cd}=l_{cd}/\xi$ implicitely via $\rsq(X_{cd},Y)\equiv
      \xi^2 X_{cd}^2/3$. The results for $X_{oc}$ presented in (b) are
      decay lengths extracted from fits of TCFs to simple
      exponentials. For the fits we used data from the segment length
      interval $X<Y,\frac12 N_{tot}/g$. However, for $Y<10$ the decay of the
      TCFs ceases to be well described by a simple exponential (see
      also Fig.~\protect\ref{fig:BACF}). Being strongly depend
      on the data range selected for the fits (data not shown, the
      values presented in (b) for $Y<10$ thus have to be taken with a
      grain of salt. We note that the
      results for datasets with different coupling constant scale
      quite well and that there is good qualitative agreement between
      the two methods.  The results nicely follow the \OSFKK
      prediction for reduced screening lengths $Y>10$, but are
      strongly influenced by excluded volume effects for smaller
      values of $Y$. In particular, the extracted persistence lengths
      systematically {\em exceed} the \OSFKK estimate.  With respect
      to finite chain length effects the first method turns out to be
      more robust than the second.  }
  \end{center}
\end{figure}

Similar conclusions can be drawn from an analysis of the
tangent-correlation function (Fig.~\ref{fig:BACF}).
In fact, our discussion in section~\ref{sec:consequences} shows that
Figs.~\ref{fig:LpCrossover} (b,d) and the log-log plots in
Figs.~\ref{fig:BACF} (a,b) are directly comparable.
Fig.~\ref{fig:BACF} (c) shows the same data in the semi-logarithmic
representation commonly used to identify a simple exponential decay of
the correlation function. Clearly, the \OSFKK scaling does not work
perfectly up to the \OSFKK persistence length, but, at least qualitatively,
we observe the expected slow-down of the decay of the correlations.

The discrepancy between our conclusion, $l_e\sim \kappa^{-y}$ with
$y=2$, and the results of previous numerical and experimental
investigations, $y\ll 2$,
can be traced back to the definition of the electrostatic persistence
length. So far we have used an indirect method (scaling plots of
segment radii and TCFs).
More direct methods usually proceed by (i) defining an {\em apparant}
electrostatic persistence length $l_{e,app}$, 
(ii) calculating $l_{e,app}$ for numerical or experimental data, 
(iii) plotting $l_{e,app}/\xi$ as a function of the reduced screening length
$Y$ (Fig.~\ref{fig:FittedLp}), and (tentatively)
(iv) extracting effective values for $y$.
Suitable definitions for $l_{e,app}$ were recently reviewed by
Ullner and Woodward~\cite{Ullner_mm_02}.
Fig.~\ref{fig:FittedLp} locates our results
for the ``orientational correlation length'' $l_{oc}$ and
the ``crossover distance'' $l_{cd}$ in our map of conformation space.
The first length is defined as the decay length of a simple exponential
fitted to the TCF while the second tries to identify the crossover
from the blob pole Eq.~(\ref{eq:r2_blobpole}) to an undulating
blob chain Eqs.~(\ref{eq:r2_SAW_BJ}) or (\ref{eq:r2_SAW_KK}).

Clearly, the extracted length scales can only be identified with
the electrostatic persistence length $l_e$ as long as 
$l_e$ is well separated from other relevant length scales,
i.e. for sufficiently large chain and screening lengths.
Fig.~\ref{fig:FittedLp} shows that our results closely follow
the $\mathrm{KK}_1$ line for $N=4096$ and $Y>10$.
However, the violation of either of the two conditions leads
to deviations. Relative to the \OSFKK prediction the
extracted persistence lengths
\begin{description}
\item[increase] for small reduced screening lengths due to excluded
volume effects and 
\item[decrease] in the opposite limit
due to finite chain length effects which are particularly strong
for the TCF and quantities related to it.
\end{description} 
Depending on the definition of $l_{e,app}$ and the range of chain and
screening lengths studied, the combination of these two effects can
lead to the observation of effective exponent $l_{e,app}\sim
\kappa^{-y}$ which are much smaller than $y=2$.  However, since the
weak $\kappa$ dependence of $l_{e,app}$ is an artifact of the
definition of the quantity, there seems to be no contradition to the
\OSFKK theory.  Attempts along these
lines~\cite{Micka_pre_96,Ullner_jcp_97,Ullner_mm_02} therefore risk
to create more confusion than insight as long as $l_{e,app}$ is not
defined within a theoretical framework which explicitly accounts for
excluded volume effects~\cite{Reed_mm_91}.

\section{Discussion}

In this paper we have combined a scaling analysis
of the conformational properties of intrinsically flexible
polyelectrolytes with Debye-H\"uckel interactions with extensive
Monte Carlo simulations of isolated chains. Our study was
focused on the controversial case of polyelectrolytes beyond
the OSF limit, i.e. to the case where the electrostatic 
screening length $\kappa^{-1}$ exceeds the bare persistence
length of the polymers in the absence of electrostatic interactions.

Our main result is the refutation of
theories~\cite{BarratJoanny_epl_93,Ha_mm_95} predicting an
electrostatic persistence length scaling as $\kappa^{-1}$. In
contrast, we have observed no significant deviations from the scenario
proposed by Khokhlov and
Khachaturian~\cite{KhokhlovKhachaturian_pol_82} who combined the idea
by de Gennes et al.~\cite{GPVB_jp_76} of a stretched chain of
polyelectrolyte blobs with the Odijk-Skolnick-Fixman theory of the
electrostatic persistence
length~\cite{Odijk_jpspp_77,SkolnickFixman_mm_77} and the
electrostatically excluded
volume~\cite{Odijk_jpspp_78,FixmanSkolnick_mm_78} between chain
segments.
Our results suggest that it is indeed possible to understand
DHWLC by considering a  hierarchy of effects due to interactions
between different classes of monomer pairs:
\begin{description}
\item[Stretching] due to the (effectively unscreened) Coulomb
repulsion between neighboring monomers into a chain
of blobs which has a finite 
\item[Bending rigidity] due to the screening of interactions
between monomers with a distance larger than $g/(\kappa\xi)$
along the chain. As a consequence, the blob chain remains straight up
the electrostatic persistence length $l_e=\kappa^{-2}/\xi$. 
Beyond $l_e$ the chain behaves like a random walk, before 
\item[Swelling] due the electrostatically excluded volume between
chain segments with a distance larger than $l_e$ becomes
relevant beyond the Flory length $l_F=\kappa^{-4}/\xi^3$
\end{description}

An interesting side result of our work is the extension of
the polyelectrolyte blob scaling to the case of strongly
interacting, almost fully stretched chains for which the
OSF theory is known to work~\cite{Odijk_jpspp_78}. 
This extension was done in the logic of the \OSFKK theory but is
incompatible with the ansatz of BJ. Its success provides strong
evidence for the irrelevance of
longitudinal and transverse fluctuations within the blob 
chain~\cite{LiWitten_mm_95}
and proves the KK idea almost by itself.

Clearly, scaling arguments cannot do justice to the full complexity of
the problem. Omitting all numerical prefactors, the ubiquituous
logarithmic corrections, finite chain length effects and, in our
opinion most importantly, a refined description of the crossovers
between narrow neighboring regimes, they cannot hope (and should not
be expected) to describe numerical or experimental data in detail.
Quite obviously, these features call for a {\em quantitative}
explanation. While our numerical results can serve as benchmarks for
the development of theories, it is a sobering thought that the {\em
  simplest} model of a {\em single, isolated} polyelectrolyte chain is
still unsolved. Compared to the much better understood neutral
polymers, the \OSFKK theory represents the equivalent of the standard
Flory argument for the excluded volume effect.  Nevertheless, we
believe to have shown that the \OSFKK theory provides the indispensible
``big picture'' needed for the design and analysis of experiments and
computer simulations.

\section{Acknowledgements}
We gratefully acknowledge helpful discussions with B. D\"unweg,
J.-F. Joanny, K. Kremer and H. Schiessel. RE is supported by
an Emmy-Noether fellowship of the DFG.

\bibliographystyle{unsrt}
%present address: Argonne National Laboratory,  Materials Science Division, Build. 212,  9700 S. Cass Avenue, Argonne, IL--60439, USA
\bibliography{../PE}

\end{document}